\documentclass[a4paper]{article}

\usepackage{INTERSPEECH2020}

\usepackage{enumitem}
\usepackage[dvipsnames]{xcolor}
\usepackage{multirow}
\usepackage{array}

\usepackage{setspace,url}
\usepackage{lineno,hyperref,multirow,cite}
\usepackage{authblk}
\usepackage{epsfig,amssymb}
\usepackage[utf8]{inputenc}
\usepackage{amsmath,graphicx}
\usepackage{amssymb,epsfig,url,mathrsfs} \usepackage{algorithm,algorithmic}
\usepackage{xcolor}
\usepackage{multirow}
\usepackage{multicol}
\usepackage{enumitem}

\usepackage{numprint}
\npthousandsep{,}

\usepackage{graphicx}
\usepackage{subfig}
\usepackage{array, makecell}
\usepackage{verbatim}
\usepackage[utf8]{inputenc}
\usepackage[T1]{fontenc}
\usepackage[english]{babel}
\usepackage{tabularx}  
\usepackage{ragged2e}  
\newcolumntype{Y}{>{\RaggedRight\arraybackslash}X} 
\usepackage{booktabs}

\title{NWPU-ASLP System for the VoicePrivacy 2022 Challenge }

\name{Jixun Yao, Qing Wang, Li Zhang, Pengcheng Guo, Yuhao Liang, Lei Xie}

\address{Audio, Speech and Langauge Processing Group (ASLP@NPU), School of Computer Science,\\Northwestern Polytechnical University, China} 
\email{\{yaojx, liangyuhao\}@mail.nwpu.edu.cn, \{qingwang, pcguo, lxie\}@nwpu-aslp.org, lizhang.aslp.npu@gmail.com}
\begin{document}

\maketitle
\begin{abstract}
This paper presents the NWPU-ASLP speaker anonymization system for VoicePrivacy 2022 Challenge. Our submission does not involve additional Automatic Speaker Verification (ASV) model or x-vector pool. Our system consists of four modules, including feature extractor, acoustic model, anonymization module, and neural vocoder. First, the feature extractor extracts the Phonetic Posteriorgram (PPG) and pitch from the input speech signal. Then, we reserve a pseudo speaker ID from a speaker look-up table (LUT), which is subsequently fed into a speaker encoder to generate the pseudo speaker embedding that is not corresponding to any real speaker. To ensure the pseudo speaker is distinguishable, we further average the randomly selected speaker embedding and weighted concatenate it with the pseudo speaker embedding to generate the anonymized speaker embedding. Finally, the acoustic model outputs the anonymized mel-spectrogram from the anonymized speaker embedding and a modified version of HifiGAN transforms the mel-spectrogram into the anonymized speech waveform. Experimental results demonstrate the effectiveness of our proposed anonymization system.
\end{abstract}
\section{Introduction}\label{sec:baseline}

Speech data on the Internet are proliferating exponentially because of the emergence of virtual assistants, telecommunication and voice pay. Such applications send recorded personal data to centralized servers where the speech data get processed and stored~\cite{meyer2022speaker}. However, speech data contain rich personal sensitive information that can be disclosed by automated systems. This includes, e.g., age, gender, health state and religious beliefs. Thus, with new regulations such as the General Data Protection Regulation (GDPR) in the EU, which strengthen privacy preservation and protection of personal data~\cite{GDPR}, how to suppress the identity information of a speaker and leave the linguistic content unchanged for telecommunication and downstream automated systems including automatic speech recognition (ASR) is becoming an urgent problem.

The hiding of speaker identity, also referred to as \textit{speaker anonymization}, is an effective technology to protect personal privacy that is becoming increasingly critical in light of the exponential growth of voice data. Most of the anonymization methods were inspired by the x-vector based anonymization~\cite{EURECOM+5909} and modified version~\cite{meyer2022speaker,gupta2020design, dubagunta2020adjustable}. The x-vector based anonymization method requires an ASV model and an x-vector pool, and simply averages different x-vectors to cancel speaker identity. To increase the privacy protection ability of x-vector based method, Mawalim \textit{et al.} used singular value decomposition (SVD) of the matrix constructed from the utterance-level speaker x-vectors~\cite{Xres}. Different from modifying utterance-level speaker x-vectors, Champion \textit{et al.} suggested F0 as the critical factor and modified it to achieve anonymization~\cite{champion2020study}. Aiming at improving the robustness of anonymization model, Gaussian Mixed Model (GMM) was adopted in~\cite{turner2020speaker} to sample x-vectors in a principal components analysis (PCA) compact space where the original distribution of cosine distances between x-vectors is retained. Note that these approaches rely on additional ASV models or x-vectors pools and increase the within-class distance of the anonymized speaker. 

Despite a lot of research in this area, speaker anonymization remains in its infancy. Consequently, the first VoicePrivacy 2020 Challenge was launched which focused on developing anonymization solutions for speech technology~\cite{tomashenko2020voiceprivacy,tomashenko2020introducing,tomashenko2022voiceprivacyresult}. In a broad view, anonymization requires altering not only the speaker’s voice, but also linguistic content, extralinguistic traits, and background sounds which might reveal the speaker’s identity~\cite{tomashenko2022voiceprivacy}. As a step towards this goal, the second edition -- VoicePrivacy 2022 challenge -- focuses on voice anonymization.

This paper aims at contributing to such anonymization systems and serves as a submission to the VoicePrivacy 2022 challenge. Different from the baseline system, our system does not involve additional ASV models or an x-vectors pool, which also reduces the risk of insufficient generalization of the ASV model and the complexity of anonymization computation. We propose an effective speaker anonymization method using two types of speaker embedding generated by a speaker encoder. Specifically, we reserve a pseudo speaker ID to generate pseudo speaker embedding ensuring the anonymized result is an artificial voice which is not corresponding to any real speaker. Furthermore, we average the randomly selected speaker embeddings, while the anonymized embedding is the weighted concatenation of averaged embedding and pseudo speaker embedding. Concatenation weight is a hyper-parameter chosen empirically. Given the linguistic feature and anonymized embedding, our method uses a neural acoustic model and a neural vocoder to generate an anonymized speech waveform. Experiments show that our proposed anonymization method leads to a notable increase in both primary metrics compared with baseline systems. Additionally, results show that the proposed anonymization system can suppress personally identifiable information and provide superior intelligibility to baseline systems.


\section{System Overview}\label{sec:baseline}
\begin{figure*}[ht]
        \centering
        \includegraphics[width=0.8\linewidth]{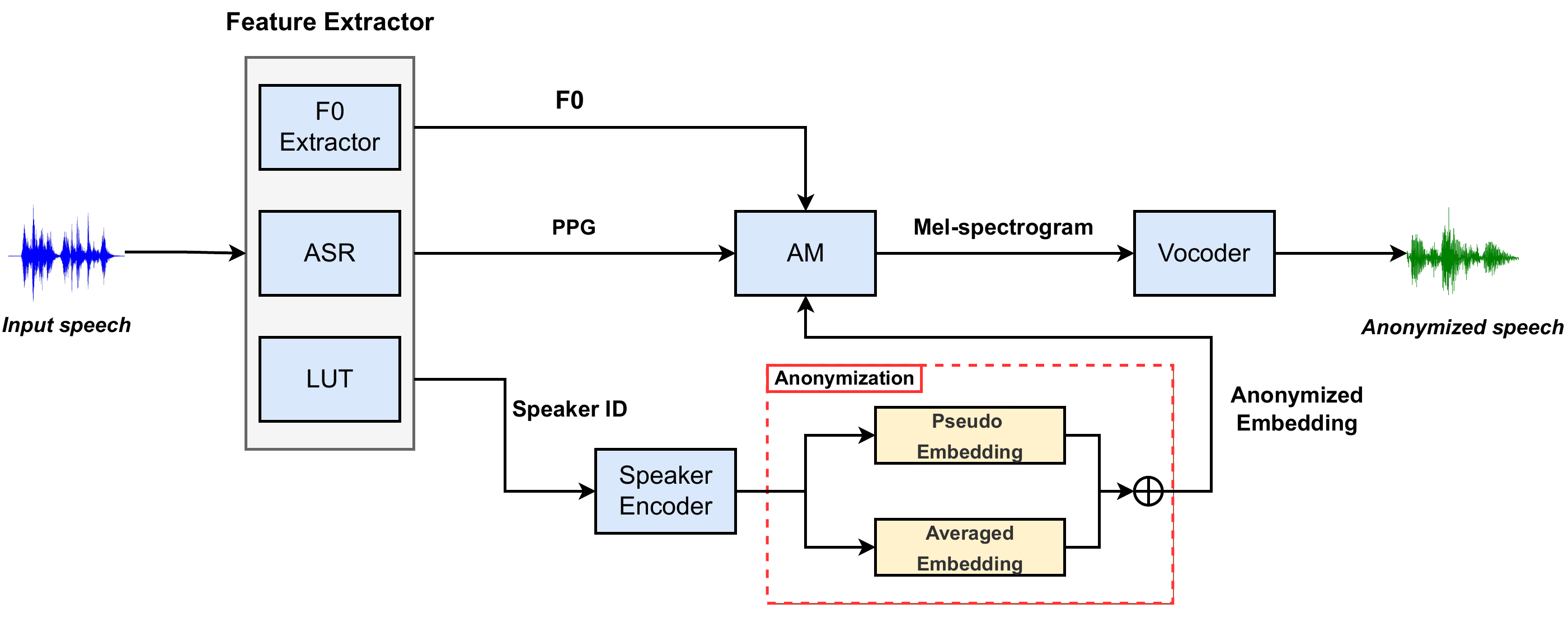}
        \vspace{-1em}
        \caption{System framework of proposed speaker anonymization system}
         \vspace{-10pt}
        \label{fig:model}\vspace{-10pt}
\end{figure*}

In this section, we will briefly introduce the framework of our proposed anonymization system which is presented in Fig.\ref{fig:model}. The anonymization system consists of four modules: (a) feature extraction, (b) acoustic model (AM), (c) anonymization module and (d) vocoder. 

As shown in Fig.1, feature extraction module extracts fundamental frequency (F0), PPG and speaker ID from input speech. During training, speaker encoder is utilized to control the speaker identity by a one-hot speaker ID. Mel-spectrum is predicted by F0, PPG and speaker embedding using the acoustic model. Finally, vocoder reconstructs the generated mel-spectrum to waveform.

In look-up table (LUT), we reserve a pseudo speaker ID to generate pseudo speaker embedding.  Based on the proposed framework, the speaker encoder described before, is used as an anonymized embedding generator. That is, the randomly selected speaker embedding as the condition for the pseudo speaker embedding that generates as output the anonymized embedding.

\section{Implementation Details}\label{sec:baseline}
\subsection{Feature extraction}
Due to the irregular periodicity of the glottal pulse, we often hear creaky voice in speech, which is usually manifested as jitter or sub-harmonics in signals~\cite{ardaillon2020gci}. This situation makes hard for F0 extractor to extract correct F0. To solve this problem, we take a hint from~\cite{choi2021neural}, which is an improvement with YIN algorithm~\cite{de2002yin}. For details, we set 80 bins of Yingram to represent a semitone range and Yingram to represent the frequency between 10.77 Hz and 768.40 Hz by setting window size to 1024.

It is crucial to extract rich linguistic information from the input speech to reconstruct an intelligible waveform and achieve a lower word error rate (WER). To this end, we resort to WeNet toolkit~\footnote{\hyperlink{}{https://github.com/wenet-e2e/wenet}}. WeNet is an open-source speech recognition toolkit in which a new two-pass approach named U2++ implemented to end-to-end (E2E) speech recognition in a single model~\cite{yao2021wenet}. U2++ consists of three parts, shared encoder, connectionist temporal classification (CTC) decoder~\cite{amodei2016deep} and attention decoder. We train U2++ on the \textit{LibriSpeech-600}\cite{panayotov2015librispeech} and \textit{LibriTTS-600}\cite{zen2019libritts},  consisting of clean-100 and other-500. At the same time, we select the top 10 models with the best WER metrics for fusion. The extracted features from the shared encoder have shown superior performance on downstream tasks such as voice conversion. Therefore, we use the shared encoder output of 512-dimensional as PPGs that can provide rich linguistic information. We generated the LUT using the LibriTTS-600 dataset, which contains 1407 real speakers and 1 pseudo speaker. The speaker embedding is generated from the original speaker ID during training and the average embedding and pseudo embedding in the LUT during inference.

\subsection{Acoustic model}
We follow the implementation of Tacotron~\cite{shen2018natural} and delete the attention layer in the CBHG module because the PPGs and the acoustic features are initially aligned. The CBHG module consists of three parts: a set of 1D convolution layers, a multi-layer highway network~\cite{srivastava2015highway} and a bidirectional gated recurrent unit (GRU) layer. Highway network and bidirectional GRU are used to extract higher-level features and contextual features of the sequence, respectively. Speaker embedding is concatenated on the final output of CBHG module as input to decoder. The decoder utilizes an autoregressive model to improve the quality and intelligibility of the reconstructed mel-spectrogram.
\begin{table}[ht]
\setlength{\abovecaptionskip}{0.5cm}
\caption{Acoustic model architectures and hyper-parameters. "k*conv-c-ReLU" denotes Conv1D with kernel k and c output channels with ReLU activation. FC means fully-connect layer}\label{tab:model}

\renewcommand\arraystretch{1.5}
  \centering
  \footnotesize
 \vspace{-15pt}
\begin{tabular}{|c|c|}

\hline
Model                         & Layer                         \\ \hline
\multirow{5}{*}{Encoder CBHG} & Conv1D:8*conv-256-ReLU        \\ \cline{2-2} 
                              & MaxPooling:stride=1,width=2   \\ \cline{2-2} 
                              & Conv1D:2*conv-256-BatchNorm1D \\ \cline{2-2} 
                              & Highway:4*fc-128-ReLu         \\ \cline{2-2} 
                              & BGRU:128 cells                \\ \hline
Decoder Pre-net               & FC:2*fc-256-ReLU-Dropout      \\ \hline
Decoder RNN                   & GRU:256 cells                 \\ \hline
Decoder Post-net              & Conv1D:4*conv-256-BatchNorm1D \\ \hline
\end{tabular}
\vspace{-10pt}
\end{table}

Both \textit{LibriTTS-clean-100} and \textit{LibriTTS-other-500} corpora\cite{zen2019libritts} are used for model training. Audio data is processed into 80-dimensional mel--spectrograms with 50ms frames and 12.5ms frame-shift. In the training stage, we use batch size of 32, Adam optimizer with learning rate of 0.001 and decay at each 10 epoch with 0.7. Models are trained for 500k steps on 1 NVIDIA GTX 2080 GPU. The detailed network architecture and hyperparameters are presented in Table
~\ref{tab:model}.

\begin{table*}[]
  \caption{Privacy results on different conditions. EER achieved by $ASV_{eval}^{anon}$ on data processed by our anonymization method vs. EER achieved by baseline B1.a or B1.b and original(Orig).}\label{tab:asv-results1}
  \centering
  \footnotesize
  \renewcommand{\tabcolsep}{0.11cm}
  \begin{tabular}{|c|c|c|c|c|c|c|c|c|c|}
\hline
\textbf{Dataset} & \textbf{Gender} & \textbf{Weight} & \textbf{Orig} & \textbf{B1.a} & \textbf{B1.b} & \textbf{Condition1} &\textbf{Condition2} &\textbf{Condition3} &\textbf{Condition4} \\ \hline \hline
\multirow{2}{*}{LibriSpeech-dev} & female & 0.25 & 8.67 & 17.76 & 19.03 & 13.92      & 21.02      & 25.28      & 26.28\\ 
                            & male & 0.25 & 1.24 & 6.37  & 5.59 & 15.53      & 19.57      & 22.05      & 23.45\\ \hline
\multirow{2}{*}{VCTK-dev (diff)} & female & 0.20 & 2.86 & 12.46 & 8.25 & 18.36      & 29.14      & 38.80      & 40.31\\ 
                            & male & 0.20 & 1.44 & 9.33  & 6.01 & 22.28      & 31.46      & 36.92      & 37.77\\ \hline 
\multirow{2}{*}{VCTK-dev (comm)} & female & 0.05 & 2.62 & 13.95 & 9.01 & 19.19      & 26.45      & 34.59      & 35.76\\ 
                            & male & 0.05 & 1.43 & 13.11 & 9.40 & 21.37      & 29.91      & 37.04      & 37.89\\ \hline 
\multicolumn{3}{|c|}{Weighted average dev}  & 3.54 & 11.74 & 9.93 & 17.51     & 25.08     & 30.55     & 31.73\\ \hline

\multirow{2}{*}{LibriSpeech-test} & female & 0.25 & 7.66 & 12.04 & 9.49 & 16.61      & 17.88      & 20.99      & 22.08\\ 
            & male & 0.25 & 1.11 & 8.91 & 7.80 & 10.69      & 14.03      & 17.37      & 19.15\\ \hline
\multirow{2}{*}{VCTK-test (diff)} & female & 0.20 & 4.89 & 16.00 & 10.91 & 23.10      & 34.83      & 40.84      & 40.64\\ 
                    & male & 0.20 & 2.07 & 10.05 & 7.52 & 23.19      & 30.20      & 37.54      & 38.81\\ \hline 
\multirow{2}{*}{VCTK-test (comm)} & female & 0.05 & 2.89 & 17.34 & 15.32 & 23.99      & 34.68      & 40.46      & 40.46\\ 
                        & male & 0.05 & 1.13 & 9.89 & 8.19 & 23.16      & 32.20      & 38.14      & 38.70\\ \hline 
\multicolumn{3}{|c|}{Weighted average test} & 3.79 & 11.81 & 9.18 & 18.44     & 24.32     & 29.19     & 30.15\\ \hline
  \end{tabular}
  \vspace{-10pt}
\end{table*}

\subsection{Anonymization Strategy}
Our anonymization strategy produces a final anonymized embedding based on the proposed framework combining averaged embedding and pseudo speaker embedding. All the kinds of embedding are explained in detail below.

\textbf{\textit{Averaged embedding}}. During training, the speaker IDs of all the training data are stored in the LUT, which is used as the speaker embedding pool. K speaker embeddings are randomly selected by the LUT, and then are averaged to get the averaged embedding. In our strategy, the averaged embedding is similar to the mean x-vector of a set of randomly selected x-vectors from an x-vector pool. But we do  not need an x-vector pool, and the LUT is equivalent to our x-vector pool.

\textbf{\textit{Pseudo speaker embedding}}. A pseudo speaker ID is reserved in the LUT, which does not correspond to any of the real speakers in the training data. This means that the pseudo speaker ID is not involved in optimizing the training process. The real speaker ID and the pseudo speaker ID produce different distributions of speaker embedding during inference.

\textbf{\textit{Anonymized embedding}}. Finally, the averaged speaker embedding and the pseudo speaker embedding weighted concatenate to generate the anonymized embedding as output. The averaged speaker embedding is in place to guarantee that each trial utterance from a particular speaker is delivered by a single pseudo-speaker, while the trial utterances from various speakers are delivered by different pseudo-speakers.
\subsection{Vocoder}

The vocoder is our modified version of HifiGAN generator~\cite{kong2020hifi}, which can effectively reconstruct high-quality waveform. Specifically, we improve the original HifiGAN by the following aspects.  First, we increase the receptive field of the generator. Second, we use the multi-resolution STFT loss to better measure the difference between fake and real speech. Last, following the setting of multi-band MelGAN~\cite{yang2021multi}, we extend HifiGAN with multi-band processing as follows. The generator takes mel-spectrograms as input and produces sub-band signals instead of full-band signals.  The modified  HifiGAN is proven to be beneficial to speech generation. As for the upsampling module in our HifiGAN, 200x upsampling is conducted through 3 upsampling layers with 2x, 5x and 5x factors respectively because of predicting 4 sub-bands simultaneously, the output channels of the 3 upsampling networks are 256, 128 and 64, respectively, and each residual dilated convolution stack has 3 layers with dilation 1, 3, 5 with kernel-size 3, 7, 11.

The modified HifiGAN vocoder is trained with batch size of 16 and AdamW optimizer with learning rate of 2e-4 on 1 NVIDIA GTX 2080 GPU. The discriminator uses the same optimization settings, with a fixed learning rate of 2e-4. As for the generator, to reduce training time and GPU memory usage, we employ a windowed generator for training, randomly sampling segments of mel-spectrogram with a window size of 50 frames as input to the vocoder. We use the same training data as the training data of the acoustic model and the ground truth alignment data to train 800k and 200k steps, respectively. 
\section{Evaluations and results}\label{sec:baseline}

\subsection{VoicePrivacy 2022 Challenge}
According to the VoicePrivacy 2022 Challenge, the anonymization systems should: (a) preserve the linguistic content and paralinguistic attributes; (b) conceal the speaker identity; (c) ensure that all trial utterances from a given speaker are uttered by the same pseudo-speaker and pseudo-speaker distinguishability~\cite{tomashenko2022voiceprivacy}. The training, development and evaluation corpora are the same as for the VoicePrivacy 2020 Challenge. \textit{VoxCeleb-1,2}~\cite{nagrani2017voxceleb} corpus and subsets of the \textit{LibriSpeech}\cite{panayotov2015librispeech} and \textit{LibriTTS}\cite{zen2019libritts} corpora are used for training voice anonymization systems. The development set comprises \textit{LibriSpeech dev-clean} and a subset of the \textit{VCTK} corpus, denoted \textit{VCTK-dev}. For \textit{VCTK-dev}, there are two trial subsets: \textit{common} and \textit{different}, \textit{common} subset is intended to support subjective evaluation of speaker verifiability in a text-dependent manner\cite{tomashenko2022voiceprivacy}. The evaluation set is similar to the development set.
\begin{table}[]
  \caption{Primary utility evaluation: WER achieved by \emph{$ASR_\text{eval}^{anon}$} on data processed by our anonymization method (with the \textrm{large  LM}). C* denotes different target EER conditions}\label{tab:asr-results1}
  \renewcommand\arraystretch{1.5}
  \centering
  \footnotesize
  \renewcommand{\tabcolsep}{0.11cm}
  \begin{tabular}{|c|c|c|c|c|c|c|c|}
\hline
\textbf{Dataset} & \textbf{Orig} & \textbf{B1.a} & \textbf{B1.b} & \textbf{C1} & \textbf{C2} & \textbf{C3} & \textbf{C4}\\ \hline \hline
LibriSpeech-dev & 3.82 & 4.34 & 4.19         & 3.91 & 3.71 & 3.65 & 3.65\\ \hline
VCTK-dev    & 10.79 & 11.54 & 10.98          & 8.10 & 7.73 & 7.68 & 7.62\\ \hline
Average dev     & 7.31 & 7.94 & 7.59         & 6.00 & 5.72 & 5.66 & 5.63\\ \hline
\hline LibriSpeech-test & 4.15 & 4.75 & 4.43         & 3.96 & 3.98 & 3.84 & 3.87\\ \hline
VCTK-test               & 12.82 & 11.82 & 10.69      & 8.37 & 7.85 & 7.81 & 7.85\\ \hline
Average test            & 8.49 & 8.29 & 7.56         & 6.16 & 5.91 & 5.82 & 5.86\\ \hline
  \end{tabular}
  \vspace{-10pt}
\end{table}

\subsection{Evaluations}
The VoicePrivacy 2022 Challenge focuses mainly on a pair of metrics: the Equal Error Rate (EER) as the privacy metric and the Word Error Rate (WER) as the primary utility metric. The basic automatic speaker verification (ASV) and automatic speech recognition (ASR) systems are trained on the \textit{LibriSpeech-train-clean-360} and retrained on the utterance-level anonymized data of \textit{LibriSpeech-train-clean-360} (denoted as $ASV_{eval}^{anon}$ and $ASR_{eval}^{anon}$). To trade-off the privacy-utility, results were divided into four conditions based on the minimum target EERs and the lower WER for a given EER condition, the better the rank of the anonymization system. The four minimum target EER condition is 15\%, 20\%, 25\% and 30\%. 

In addition to the primary metrics, another two secondary utility metrics, namely pitch correlation $\rho F0$ and the gain of voice distinctiveness ($G_{VD}$)~\cite{noe2020speech} which measured how well anonymization preserves the intonation of the original utterance and evaluated the requirement to preserve voice distinctiveness, respectively. The gain of voice distinctiveness metric is defined as the ratio of diagonal dominance of the original matrices and the anonymized matrices. These two metrics are not used for final ranking, but the anonymized results should achieve a minimum average pitch correlation of $\rho F0 > 0.3$ for each dataset and each condition. 

\subsection{Results}
The EER results are computed by averaging three EERs on \textit{LibriSpeech-test-clean}, \textit{VCTK-test (common)} and \textit{VCTK-test (different)} datasets with weights of 0.5, 0.1 and 0.4, respectively. And the WER results are computed by averaging two WERs on \textit{LibriSpeech-test-clean} and \textit{VCTK-test} with equal weights. We meet the different EER conditions by adjusting the weight of pseudo speaker embedding and averaged embedding. The weights of the pseudo speaker embedding corresponding to different conditions are 0.6, 0.8, 0.9 and 0.95, respectively, and the averaged speaker embedding weights are 0.4, 0.2, 0.1 and 0.05,  respectively.

\begin{table}[htb]
  \caption{Secondary utility evaluation: pitch correlation $\rho F0$ achieved on data processed by B1.a, B1.b and our anonymized results.}\label{tab:f0-results}
  \renewcommand\arraystretch{1.2}
  \renewcommand{\tabcolsep}{0.14cm}
  \centering
  \footnotesize
  \vspace{-10pt}
  \resizebox{0.46\textwidth}{!}{
  \begin{tabular}{|c|c|c|c|c|c|c|c|c|}
\hline
\textbf{Dataset} & \textbf{Gender} & \textbf{B1.a} & \textbf{B1.b} & \textbf{C1} & \textbf{C2} & \textbf{C3} & \textbf{C4}\\ \hline \hline
\multirow{2}{*}{LibriSpeech-dev} & female & 0.77 & 0.84 & 0.70       & 0.71       & 0.71       & 0.71\\
                                & male & 0.73 & 0.76 & 0.69       & 0.69       & 0.69       & 0.69\\ \hline
\multirow{2}{*}{VCTK-dev (dif)} & female & 0.84 & 0.87 & 0.76       & 0.76       & 0.77       & 0.76\\
                                & male & 0.78 & 0.76 & 0.71       & 0.71       & 0.71       & 0.71\\ \hline
\multirow{2}{*}{VCTK-dev (com)} & female & 0.79 & 0.84& 0.71       & 0.71       & 0.72       & 0.71\\
                                & male & 0.72 & 0.72& 0.67       & 0.67       & 0.67       & 0.67\\ \hline
\multicolumn{2}{|c|}{Weighted average dev} & 0.77 & 0.80 & 0.71       & 0.71       & 0.72       & 0.72\\ \hline

\multirow{2}{*}{LibriSpeech-test} & female & 0.77 & 0.85 & 0.71       & 0.72       & 0.72       & 0.72\\
                                & male & 0.69 & 0.72 & 0.64       & 0.64       & 0.64       & 0.64\\ \hline
\multirow{2}{*}{VCTK-test (dif)} & female & 0.87 & 0.87 & 0.77       & 0.76       & 0.77       & 0.77\\
                                & male & 0.79 & 0.77 & 0.71       & 0.71       & 0.71       & 0.71\\ \hline
\multirow{2}{*}{VCTK-test (com)} & female & 0.79 & 0.85 & 0.72       & 0.71       & 0.72       & 0.72\\
                                & male & 0.70 & 0.71 & 0.64       & 0.65       & 0.65       & 0.65\\ \hline
\multicolumn{2}{|c|}{Weighted average test} & 0.77 & 0.80 & 0.70       & 0.70       & 0.71       & 0.70\\ \hline

  \end{tabular}}
  \vspace{-10pt}
\end{table}

\subsubsection{Primary privacy results}
Privacy results on different conditions are shown in Table~\ref{tab:asv-results1}. The original speech and two baseline results are denoted as Orig, B1.a and B1.b, respectively. It is observed that our anonymization method meets all the conditions and achieves 18.44\%,24.32\%,29.19\% and 30.15\% average EERs. Our approach leads to a notable increase in average EER of up to 18.34\% compared with B1.a and 20.97\% compared with B1.b which demonstrates our anonymization method can protect speaker privacy effectively. We find that our results on \textit{Librispeech}\cite{panayotov2015librispeech} corpus for different genders perform similarly instead of the significant difference in baseline systems.

\subsubsection{Primary utility results}
As shown in Table~\ref{tab:asr-results1}, we observe that the WERs for all conditions are better than both baseline systems, with a slight improvement for the \textit{LibriSpeech-test}, and a significant improvement for the \textit{VCTK-test}. The lowest WERs are 3.84\% and 7.781\% on the \textit{Librispeech-test} and the \textit{VCTK-test}, respectively. The average WER is 2.47\% and 1.74\% lower than B1.a and B1.b which demonstrates our anonymization method can achieve better intelligibility than baseline systems.

\begin{table}[]
  \caption{Secondary utility evaluation: gain of voice distinctiveness $G_{VD}$ achieved on data processed by B1.a, B1.b and our anonymized results.}\label{tab:gvd-results}
  \renewcommand\arraystretch{1.4}
  \renewcommand{\tabcolsep}{0.14cm}
  \centering
  \footnotesize
  \resizebox{0.46\textwidth}{!}{
  \begin{tabular}{|c|c|c|c|c|c|c|c|c|}
\hline
\textbf{Dataset} & \textbf{Gender} & \textbf{B1.a} & \textbf{B1.b} & \textbf{C1} & \textbf{C2} & \textbf{C3} & \textbf{C4}\\ \hline \hline
\multirow{2}{*}{LibriSpeech-dev} & female & -9.15 & -4.92 & -2.94      & -10.50     & -17.47     & -21.35\\
                                & male & -8.94 & -6.38 & -2.69      & -9.18      & -15.78     & -18.66\\ \hline
\multirow{2}{*}{VCTK-dev (dif)} & female & -8.82 & -5.94 & -2.38      & -8.33      & -12.19     & -13.96\\
                                & male & -12.61 & -9.38 & -3.10      & -10.68     & -17.25     & -20.72\\ \hline
\multirow{2}{*}{VCTK-dev (com)} & female & -7.56 & -4.17 & -1.98      & -6.72      & -13.33     & -17.18\\
                                & male & -10.37 & -6.99 & -2.06      & -7.70      & -14.81     & -19.71\\ \hline
\multicolumn{2}{|c|}{Weighted average dev} & -9.71 & -6.44 & -2.71      & -9.442     & -15.61     & -18.86\\ \hline

\multirow{2}{*}{LibriSpeech-test} & female & -10.04 & -5.00 & -2.72      & -9.21      & -16.44     & -20.13\\
                                & male & -9.01 & -6.64 & -1.64      & -7.36      & -13.90     & -17.83\\ \hline
\multirow{2}{*}{VCTK-test (dif)} & female & -10.29 & -6.09 & -2.82      & -9.18      & -15.41     & -17.86\\
                                & male & -11.69 & -8.64 & -3.85      & -10.77     & -15.82     & -17.65\\ \hline
\multirow{2}{*}{VCTK-test (com)} & female & -9.31 & -5.10 & -2.15      & -8.12      & -15.55     & -20.39\\
                                & male & -10.43 & -6.50 & -2.68      & -9.43      & -16.78     & -21.26\\ \hline
\multicolumn{2}{|c|}{Weighted average test} & -10.15 & -6.44 & -2.67      & -9.012     & -15.46     & -18.69\\ \hline

  \end{tabular}}
\end{table}

\subsubsection{Secondary primary results}
We report in Table~\ref{tab:f0-results} and Table~\ref{tab:gvd-results} the results for the two secondary primary results. It can be clearly found that the pitch correlation of our anonymized data exceeds the minimum threshold and achieve similar results with baseline systems for each condition. The highest pitch correlation is achieved in condition 1 of 0.7.  The results of $G_{VD}$ show that voice distinctiveness worsens as the EER rises. This may be due to the increasing proportion of pseudo embedding in the anonymized embedding. Similarity matrices of pseudo speaker embedding do not have a clearly dominant diagonal, so the $D_{diag}(M_{pp})$ metric described in~\cite{noe2020speech} will converge to 0.

\section{Conclusion}\label{sec:baseline}
This paper presents our anonymization system for VoicePrivacy 2022 Challenge. The proposed system utilizes averaged embedding as the condition for pseudo speaker embedding to produce the final anonymized embedding, with the anonymized mel-spectrogram generated by the acoustic model. Finally, a ground truth alignment data retrained vocoder converts the anonymized mel-spectrograms to waveform.  Under the VoicePrivacy 2022 Challenge evaluation plan, the experiment results show that our system achieves the best average EER and WER of 30.15\% and 5.82\%, respectively. The average EER and WER results demonstrate that our anonymization method can suppress personally identifiable information in the speech signal and outperform baseline systems in terms of intelligibility.

\bibliographystyle{IEEEtran}
\bibliography{mybib}

\end{document}